# Coexistence of flat bands and Dirac bands in a carbon-Kagome-lattice family

Chengyong Zhong[1], Yuee Xie[1*], Yuanping Chen[1*] and Shengbai Zhang[2]

[1]Department of Physics, Xiangtan University, Xiangtan, 411105, Hunan, China
[2]Department of Physics, Applied Physics, and Astronomy Rensselaer Polytechnic Institute, Troy, New York, 12180, USA

**Abstract:** The Dirac bands and flat bands are difficult to coexist because they represent two extreme ends of electronic properties. However, in this paper, we propose a carbon-Kagome-lattice (CKL) family based on first-principles calculations, and the coexistence of Dirac and flat bands are observed in this series of three-dimensional carbon structures. The flat bands are originated from the orbital interactions of the Kagome lattices, while the Dirac bands are related to the carbon zigzag chains. A tight-binding model is used to explain the various band structures in different CKLs. The coexistence of flat and Dirac bands around the Fermi level implies that CKL structures maybe can serve as superconductors. In addition, electronic properties of the thinnest CKL slabs, only consisting of benzene rings, are studied. Flat bands are found in the band spectra of the two-dimensional structures, and split into spin-up and spin-down bands because of strong correlated effect in the case of hole doping.

*Corresponding author. E-mail address: xieyech@xtu.edu.cn (Y. Xie); chenyp@xtu.edu.cn (Y.Chen).

# 1. Introduction

The electronic properties of a crystalline solid can be described by energy band structure [1, 2]. Electrons move in the crystal just like free electrons except for a different effective mass as defined by the band dispersion[2]. In conventional bands, the masses of electrons are finite. However, in some special cases, the band dispersions can drastically differ from the conventional ones, which always implies new physics and new applications and thus attracts much attention [3-9].

Dirac band is one of the special energy bands[10].Its linear dispersion means the electrons are massless, which will result in very high electron mobility[10, 11]. The special bands have been found in many lattices and materials up till now [12],in which the most famous are carbon materials consisting of zigzag chains, from one dimensional (1D) (zigzag nanoribbons) to two dimensional (2D) (graphene) and to three dimensional (3D) (carbon foam)[11, 13-15]. Interesting physical phenomena associated with Dirac bands are surfaced gradually, such as quantum Hall effect and surface topological effects[16-18]

Flat band is another kind of special energy band, in which the electron effective mass should be arbitrarily large according to the electronic band theory. Trivial flat bands can be obtained easily, for example, induced by the impurities in the crystal [19-21]. While nontrivial flat bands are originated from the destructive interference of wavefunctions in the lattices [22-24], therefore, the flatness of the nontrivial bands is very sensitive to the lattice patterns and orbital interactions in real atomic structures. The Kagome and triangular lattices are two typical lattices may induce flat bands because of their geometric frustration [8, 25, 26]. Since the kinetic energy of electron is quenched in the flat band, the Coulomb interaction becomes critical, giving rise to various exotic many-body states, such as ferromagnetism, superconductivity and Wigner crystal[22, 27-30].

The Dirac bands and (nontrivial) flat bands represent two extreme ends of energy bands, like two sides of coins. Therefore, they are seemingly difficult to coexist in one band structure, especially around the Fermi level. However, in this paper, we find that the two special bands can coexist in a carbon-Kagome-lattice (CKL) family we proposed. Moreover, the energy bands around the Fermi level form so-called flat/steep bands[31, 32], which may spawn the superconductivity because of high density of states on the Fermi level. Additionally, the thinnest slabs of the CKL structures, only consisting of benzene rings, are also studied. Flat bands and spin splitting induced by strong correlated effect are observed.

## 2. Model and Computational Methods

Figure 1(a) is a sketch view of Kagome lattice, while Figs. 1(b-e) show the atomic structures of the CKL family we proposed. By comparing Fig. 1(a) and Figs. 1(b-e), one can find that the structures corresponding to the Kagome lattice as each lattice is replaced by a zigzag-edged graphene nanoribbon. The structure in Fig. 1(b) is called CKL1 as the nanoribbon only includes a zigzag chain, i.e., the width of the nanoribbon is equal to $n = 1$. In turn, as the width of the nanoribbons $n = 2, 3$, and $4$, the structures in Figs. 1(c-e) are called CKL2, CKL3 and CKL4, respectively. It should be note that the space groups of the CKL family have two classes depending on $n$ is odd or not. If $n$ is odd, the CKLn belong to $P6_3/mmc$, otherwise$P6/mmm$. The optimized lattice parameters for CKL1-CKL4 are $a$ = 4.46Å, 8.04Å, 11.83Å and 15.53Å, and $c$ = 2.53Å, 2.48Å, 2.50Å, and 2.49Å, respectively. The coordinates for each structure are presented in Table S1 in the supporting information (SI). Our previous study showed that CKL1 is stable with a cohesive energy of -8.81 eV/atom[33]. The calculations indicate that the cohesive energies ($Ec$) of CKL2 (-8.89 eV/atom), CKL3 (-8.97 eV/atom) and CKL4 (-9.01 eV/atom) are lower than that of CKL1 step by step. A comparison of *Ec* in Table 1 shows that, although the *Ec* of CKLn are high than graphite and diamond, they can be comparable with that of $C_{60}$ and are lower than those of other carbon allotropes such as Bcc-C8[34, 35], T6[36]. Therefore, these structures should be also meta stable carbon allotropes, which maybe can be fabricated by graphene nanoribbons through bottom-up self-assembly methods[37].

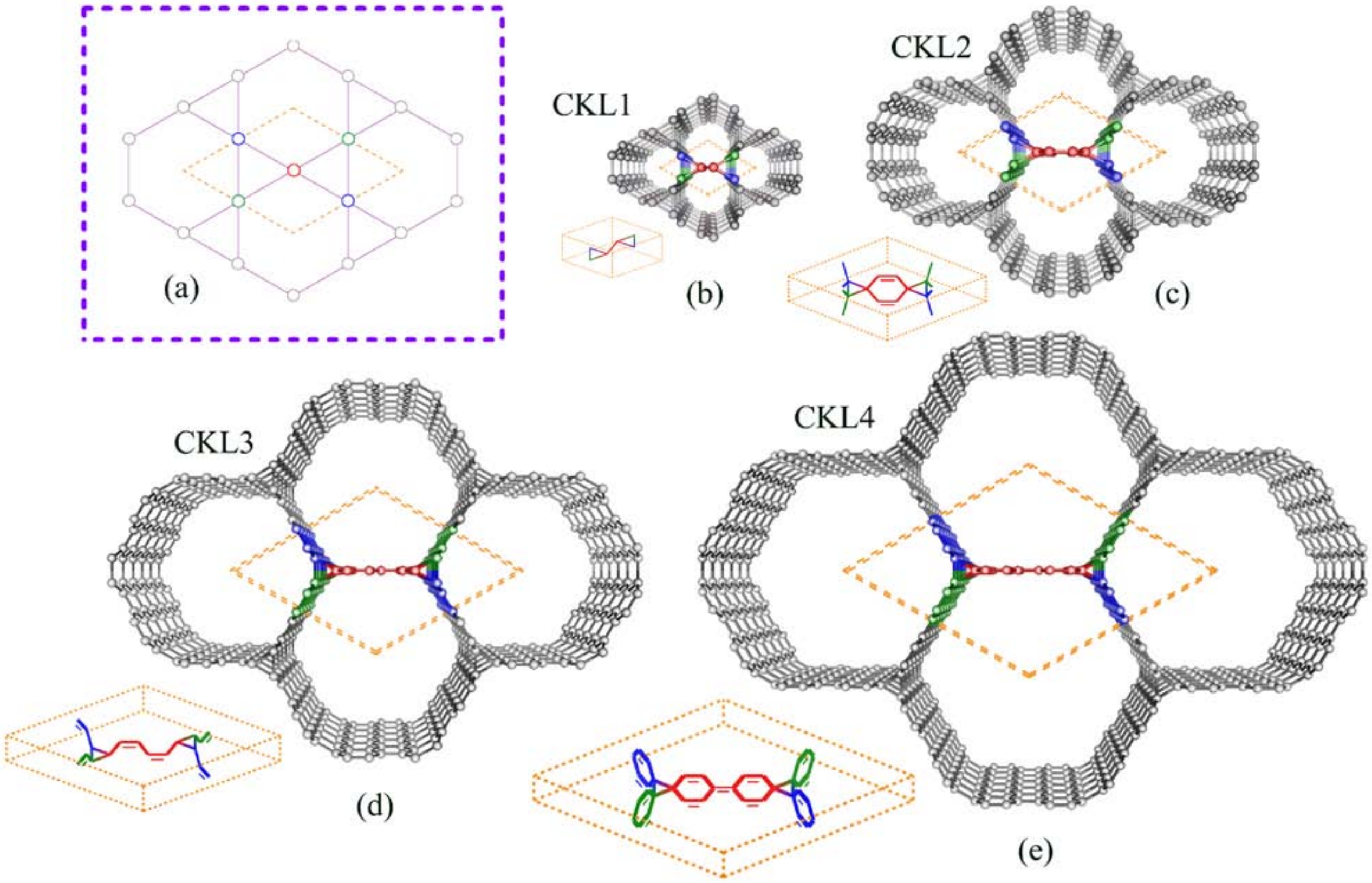

Figure 1 (a) Sketch view of a Kagome lattice. (b)-(e) Perspective views of atomic structures for the CKL family from CKL1 to CKL4, respectively, along with a side view of unit cell for each structure. CKLn (*n = 1, 2, 3* and *4*) consist of zigzag nanoribbons linked by triangular rings, where the width of each nanoribbon is *n*, i.e., the nanoribbon includes *n* zigzag chains. As each nanoribbon is condensed to a hollow point, all the structures in (b)-(e) become the Kagome lattice in (a). The diamond-shaped dashed lines represent the unit cells of the lattice and structures.

| | CKL1 | CKL2 | CKL3 | CKL4 | Diamond | Graphite | $C_{60}$ | Bcc C8[34] | T6[36] |
|---|---|---|---|---|---|---|---|---|---|
| *Ec* (eV) | -8.81 | -8.89 | -8.97 | -9.01 | -9.09 | -9.22 | -8.85 | -8.49 | -8.63 |

Table 1. The *Ec* of CKL1-CKL4, Diamond, Graphite, C60, Bcc C8 and T6.

We performed first-principles calculations within the density functional theory (DFT) formalism as implemented in VASP code[38]. The potential of the core electrons and the exchange-correlation interaction between the valence electrons are described, respectively, by the projector augmented wave[39] and the generalized gradient approximation (GGA) of Perdew-Burke-Ernzerhof (PBE)[40, 41]. The kinetic energy cutoff of 500 eV is employed. The atomic positions were optimized using the conjugate gradient method, the energy convergence value between two consecutive steps was chosen as $10^{-5}$ eV. A maximum force of $10^{-2}$ eV/Å was allowed on each atom. $9\times9\times16$, $6\times6\times16$, $4\times4\times16$ and $3\times3\times16$ were used to sample the Brillion Zone (BZ) of CKL1, CKL2, CKL3 and CKL4 with the Gamma-centered k-point sampling scheme, respectively. In the calculation of slabs, the k-point sampling of $6\times6\times1$ was used.

## 3. Results and Discussions

In Figure 2, the band structures for the CKL1-CKL4 are shown. We focus on their in-plane electronic properties along *k* path $\Gamma-M-K-\Gamma$ (Kagome-lattice plane in real space) and out-of-plane properties along $\Gamma-A$ (zigzag chain direction in real space). One can find that the energy band structures change drastically with the structures, especially from CKL1 to CKL3.To clearly show the bands evolution with the variation of the structures, red and blue curves highlight the in-plane and out-of-plane bands around the Fermi level. Fig. 2(a) illustrates that CKL1 is a direct-band-gap semiconductor. Both the in-plane and out-of-plane bands have large dispersions, indicating that the electrons have small effective masses in every direction. Fig. 2(b) shows that, in the band structure of CKL2,there exists a flat band around the Fermi level, which is a part of one Kagome bands [30].The red bands in Fig. 2(a) evolve into two Kagome bands in Fig. 2(b). Meanwhile, in the out-of-plane direction, the upper blue band shifts down and closes to the other one. Fig. 2(c) shows the band structure of CKL3, where the flat band around the Fermi level in Fig. 2(b) moves to the upper Kagome band and forms a ruby band[42]. In the out-of-plane direction, a Dirac point crossed by the two blue bands appears, and one of the bands crosses the Fermi level. In this case, a flat band and a Dirac band coexist at the Fermi level. This is similar to the coexistence of flat band and steep band studied in the former literatures[31, 32],which have discussed the possible applications for the superconductors. Therefore, the CKL3 maybe also is a superconductor material. The band structure of CKL4 is somewhat similar to that of CKL3, as shown in Fig. 2(d), where the flat and Dirac bands coexist at the Fermi level and the flat band becomes more flat nearly like a line.

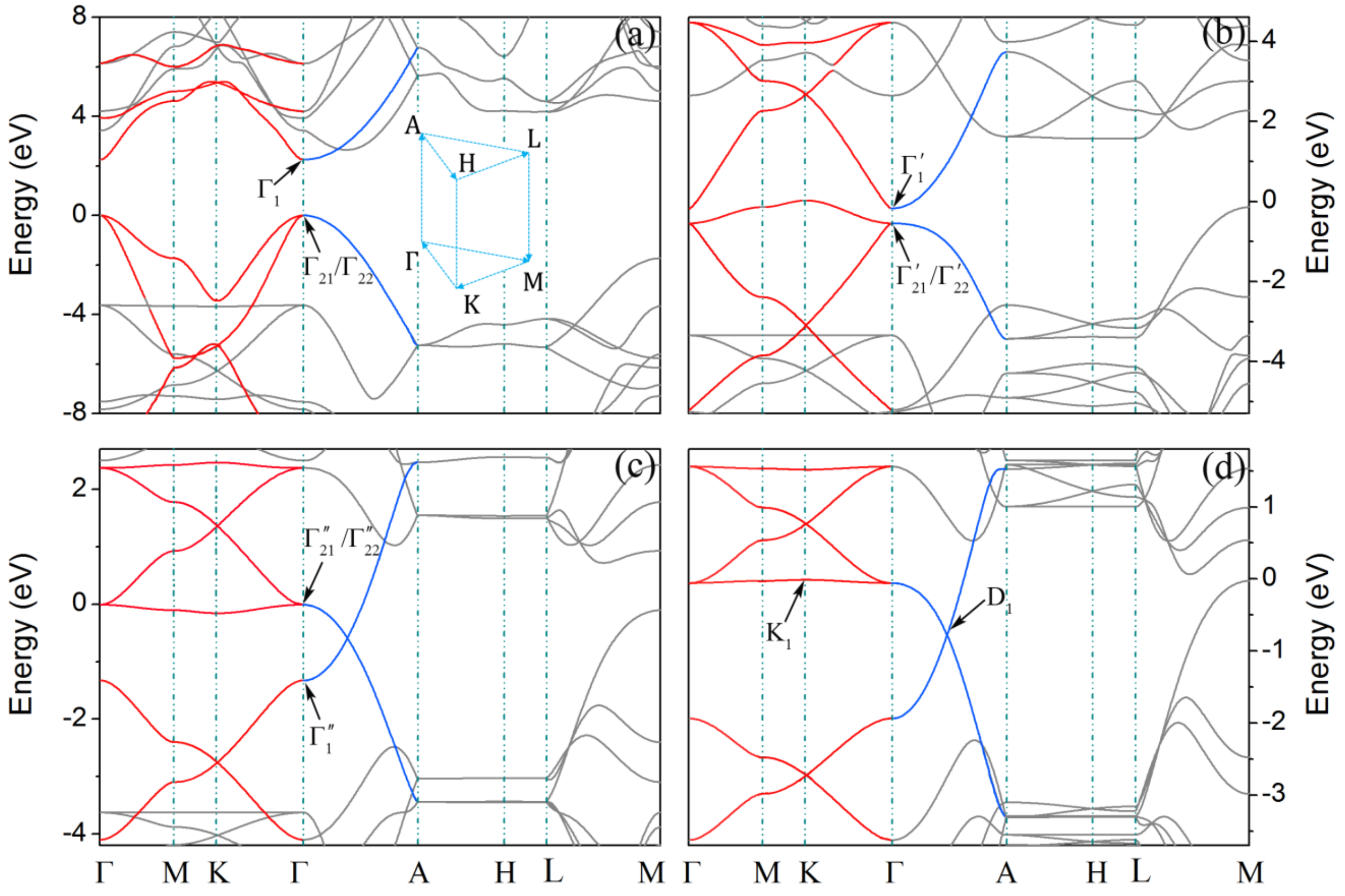

Figure 2. The electronic band structures for (a) CKL1, (b) CKL2, (c) CKL3 and (d) CKL4, respectively. The inset in (a) shows the high-symmetry k points in BZ. The labels $\Gamma_1$ and$\Gamma_{21}/\Gamma_{22}$ in (a) represent three orbital frustration states in CKL1, while $\Gamma'_1$, $\Gamma'_{21}/\Gamma'_{22}$ in (b) and $\Gamma''_1$,$\Gamma''_{21}/\Gamma''_{22}$ in (c) represent three orbital frustration states in CKL2 and CKL3, respectively. $K_1$ and $D_1$ in (d) label two states in the flatband and near the Dirac point, respectively.

To analyze the electronic properties of the CKL family, in Figs. 3(a) and 3(b), the partial density of states (PDOS) of CKL2 and CKL3 are shown respectively. It is found that the PDOS around the Fermi level (-2~2 eV) are mainly attributed by the electrons of $p_x$ and $p_y$ orbitals, while the attribution of s and $p_z$ orbital can be omitted. Note that the plane xy is perpendicular to the zigzag chains, and thus the $p_x$ and $p_y$ orbitals here actually are π orbitals of the zigzag-edged nanoribbons.

Figs. 3(c) and 3(d) present charge densities of states near the Fermi level in the CKL2 and CKL3,respectively, which clearly show that the π orbitals form π bonds inner the nanoribbons and banana-shaped σ bonds inner the triangle rings.

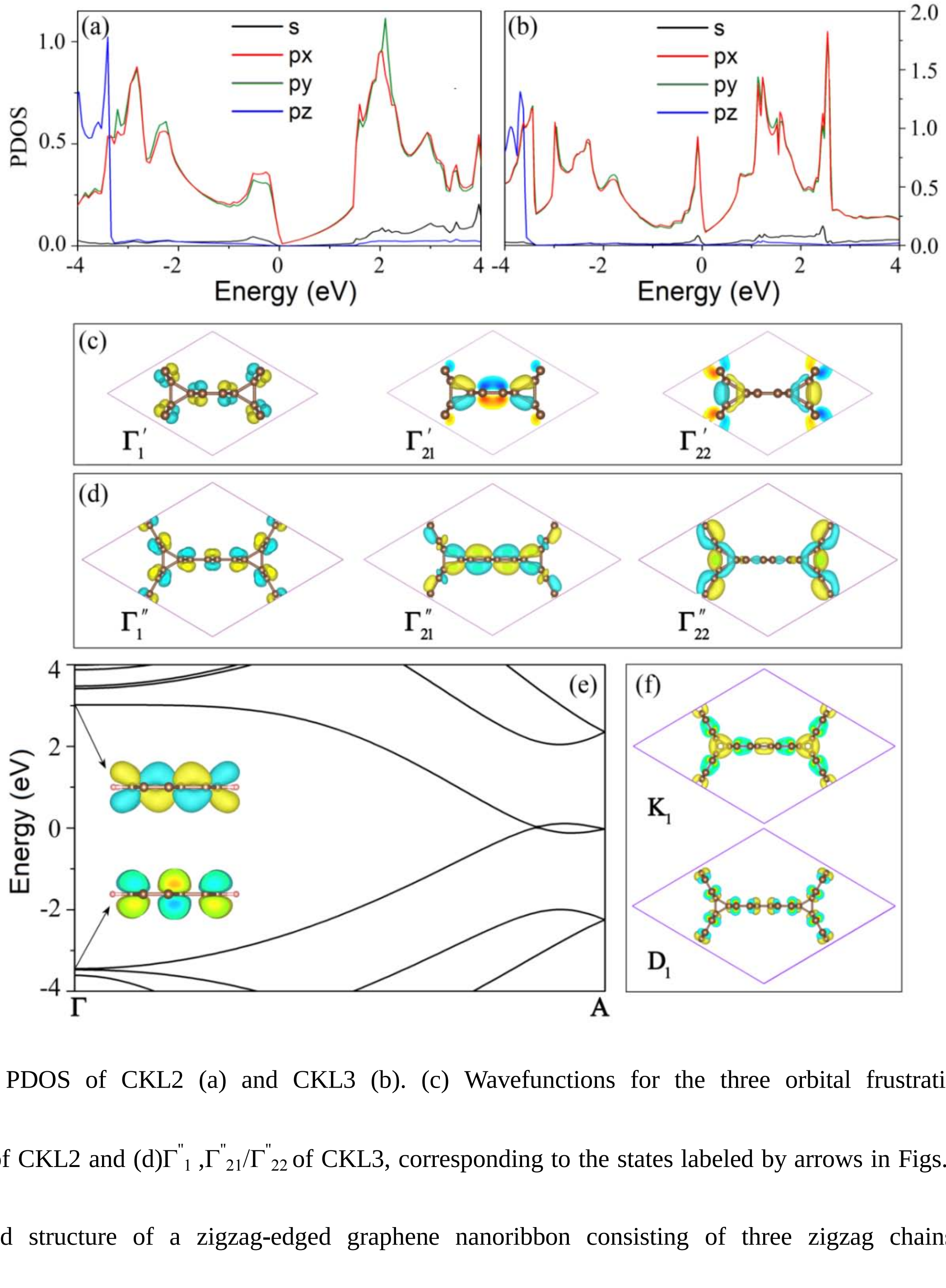

Figure 3. PDOS of CKL2 (a) and CKL3 (b). (c) Wavefunctions for the three orbital frustration states $\Gamma'_1$, $\Gamma'_{21}/\Gamma'_{22}$ of CKL2 and (d)$\Gamma''_1$ ,$\Gamma''_{21}/\Gamma''_{22}$ of CKL3, corresponding to the states labeled by arrows in Figs. 2(b) and 2(c).(e) Band structure of a zigzag-edged graphene nanoribbon consisting of three zigzag chains. Insets: Wavefunctions for the two states at Γ point labeled by the arrows. (f) The charge densities for the two states $K_1$ and $D_1$ of CKL4, corresponding to the states labeled by arrows in Fig 2(d).

Because the band edges of the CKL structures are only dominated by π orbitals, it is convenient to explain the relation between the structures and their electronic properties. First, we explain the band evolution in the out-of-plane direction, i.e., along the *k* path Γ–A. As discussed in our former study [33], there exists orbital frustration in the CKL1, i.e., a p orbital cannot find an orientation which simultaneously favors all the orbital-orbital interactions with its neighbors. The orbital frustration leads to three frustration states [33], a singlet state at $\Gamma_1$ and two degenerated states at $\Gamma_{21}/\Gamma_{22}$ (see Fig. 2(a)). The orbital frustration leads to three frustration states [33], a singlet state at $\Gamma_1$ and two degenerated states at $\Gamma_{21}/\Gamma_{22}$ (see Fig. 2(a)). As a comparison, in Figs. 3(c) and 3(d), we plot the wavefunctions of the three states $\Gamma'_1$, $\Gamma'_{21}/\Gamma'_{22}$ in Fig. 2(b) and $\Gamma''_1$ ,$\Gamma''_{21}/\Gamma''_{22}$ in Fig. 2(c), respectively. Seen from Figs. 3(c) and 3(d), the orbital frustration exists in all the structures of CKL family. In the three frustration states, $\Gamma'_{21}/\Gamma''_{21}$ or $\Gamma'_{22}/\Gamma''_{22}$ are the degenerated states while $\Gamma'_1$ or $\Gamma''_1$ is the singlet. The energies of these states will vary with structures. In the CKL1 and CKL2, the energy of the singlet is higher than that of the doublet, because in the triangular ring the former is an anti-bonding state while the latter are mixed (bonding and anti-bonding) states. Therefore, there has an energy gap along Γ–A. In CKL3 and CKL4, these states evolve into the states similar to those in a zigzag nanoribbon gradually, as seen the states in Fig. 4(e), which results in the energy of the singlet decreases while that of the doublet increases. Therefore, the two blue bands cross together and form a Dirac point. To clearly show the difference of states in the flat band and Dirac band, in Fig. 3(f), we plot two states corresponding to $K_1$ and $D_1$ in Fig. 2(d). In the state $K_1$, the electrons are tightly confined on the closed triangular ring or form π-bond dimers in the nanoribbon, indicating that there exist strong interactions between Kagome-lattice sites. In the state $D_1$, however, the electrons form π-bonds along the zigzag chains, indicating that Dirac electron channels are constructed along the chain.

Next, a tight-binding model based on the π orbitals can be used to explain the bands evolution along the $k$ path Γ–M–K–Γ. To simplify the orbital interaction between all atoms, we use a simplified model to represent the unit cells of all CKL structures, as shown in the inset of Fig. 4(a), where the dimmer lattice sites (two lattices in the middle of the unit cell) represent the zigzag nanoribbon in atomic structures. Then, the tight-binding Hamiltonian of the model is,

$$\mathrm{H} = \sum_{i,j} t_{\alpha} a_i^{\dagger} a_j , \quad (1)$$

Where $a_i^{\dagger}$ and $a_j$ represent the creation and the annihilation operators, respectively, at site $i$ or $j$, and $t_{\alpha}$ represents the hopping energy between sites. In detail, we use$t_0$ and $t_1$ to describe the nearest-neighbor hopping energies inter- and intra-nanoribbons, $t_2$,$t_3$ and $t_4$ represent the high-order next-neighbor hopping energies, as shown in the inset of Fig. 4(a). Figs. 4(a-d) show the tight-binding band structures for the CKL family and the corresponding parameters of the hopping energies are listed in Table 2. By comparing Fig. 4 and Fig. 2, one can find that the tight-binding results fit very well to the first-principles results. From CKL1 to CKL4, $t_0$ increases because the interactions between the nanoribbons are strengthened as the nanoribbons become wider; while $t_1$-$t_4$ all decrease because the wider nanoribbons result in the distances of these orbital interactions become larger. The variation of the parameters indicates that, in order to obtain flat bands in the Kagome lattice, $t_0$ should be large while other hopping energies should be small. This is consistent with the previous results which demonstrate that the flatness of the flat bands is dependent on the ratio of nearest-neighbor and next-neighbor parameters [8, 30]. To the best of our knowledge, the CKL family is the first example to reveal the variation process from conventional bands to flat bands.

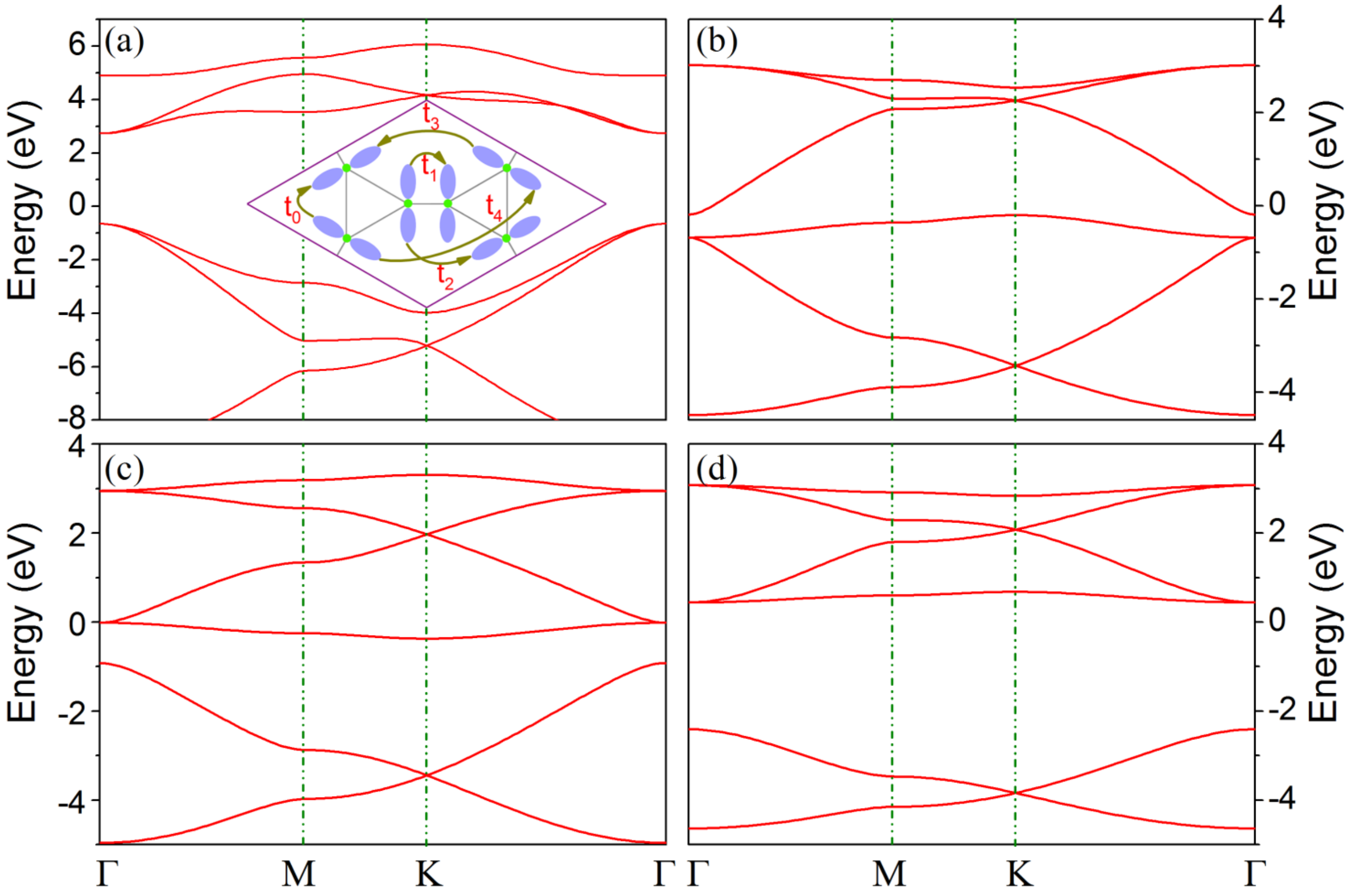

Figure 4(a)-(d) The tight-binding band structures for CKL1 to CKL4, respectively. Inset of (a): A simplified model to represent the unit cells of all CKL structures, where the blue lobes represent the p-orbitals in CKL family and $t_0$-$t_4$ represent the hopping energies between sties.

| | CKL1 | CKL2 | CKL3 | CKL4 |
|---|---|---|---|---|
| $t_0$ | -1.04 | -1.17 | -1.47 | -1.76 |
| $t_1$ | -4.23 | -1.95 | -1.66 | -1.25 |
| $t_2$ | -0.52 | -0.13 | -0.03 | -0.006 |
| $t_3$ | -0.72 | 0.16 | -0.12 | 0.08 |
| $t_4$ | 0.39 | 0 | 0 | 0 |

Table 2. The values of hopping energies $t_0$-$t_4$ in CKLn.

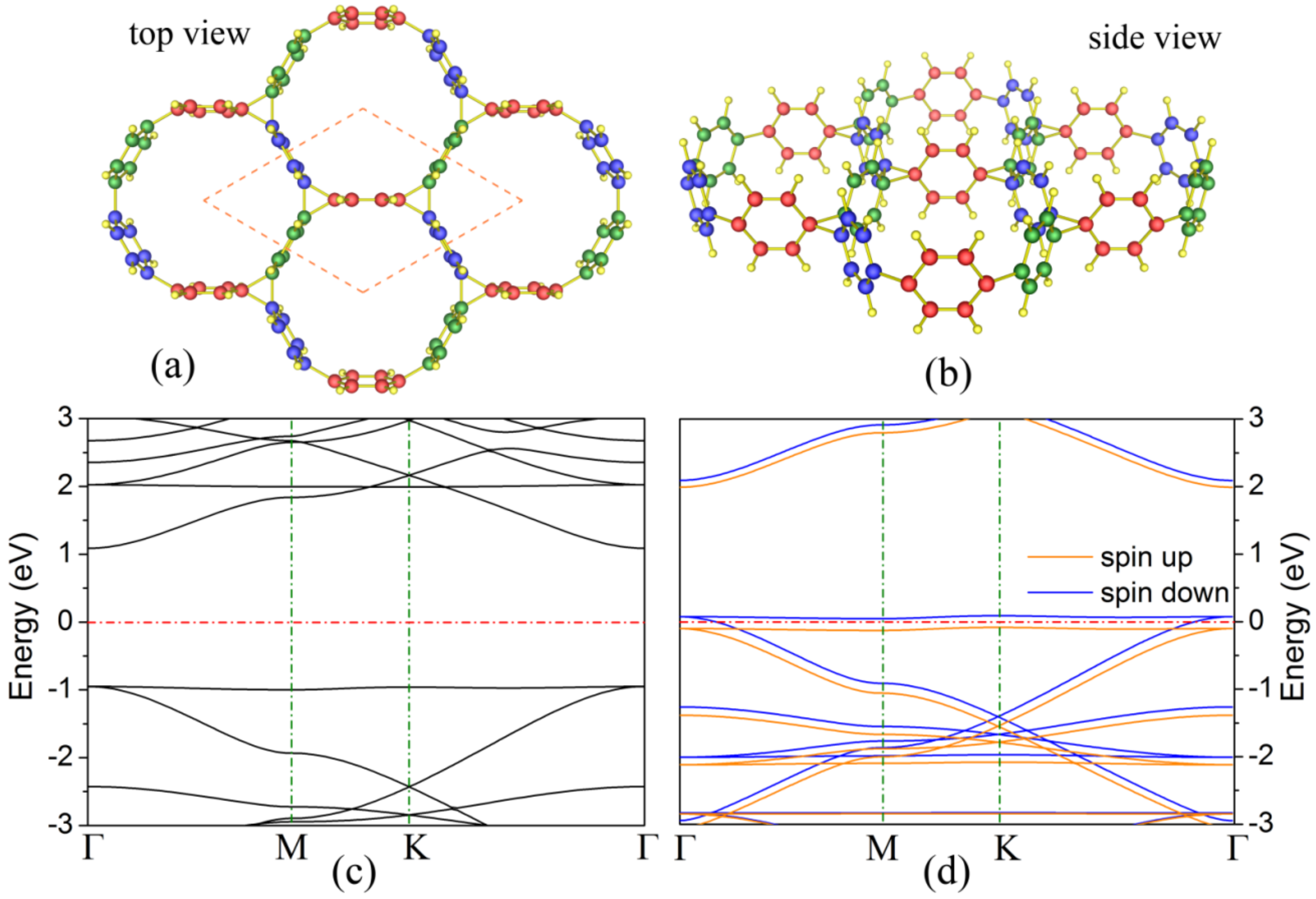

Figure 5 (a) Top view and (b) side view of the thinnest slab of CKL2.The red, blue, and green hexagons represent benzene rings, which can be viewed as sites of Kagome lattice; the yellow atoms indicate hydrogen. (b) The band structure of the slab. (c) The band structure of the slab in the case of hole doping. The red dashed lines stand for the Fermi level.

In light of the flat bands in the 3D Kagome structures along Γ–M–K–Γ, one may consider if the flat bands still retain when the 3D structures are cut into slabs. We take CKL2 as an example to study the properties of the slabs. Figs. 5(a) and 5(b) show top and side views of a thinnest slab of CKL2, respectively, which only consists of benzene rings. Graphene consists of benzene rings in a plane, while the benzene rings in the slab are perpendicular to the plane(the coordinates of the slab are provided in the SI). The slab is also a Kagome lattice where each lattice corresponds to a benzene ring. The surfaces of the slab are passivated by hydrogen atoms to make the structure more stable. Fig. 5(c) shows the band structure of the slab. There still exists flat band below the Fermi level; moreover, the band is flat than that in the band structure of CKL2. Fig. 5(d) shows the band structures for the slab in the case of hole doping, i.e., minus an electron from a unit cell of the structure[7, 9]. In experiments, the doping effect can be achieved by the electrostatic gating [9, 43]. After doping, the spin-up and spin-down bands are separated with a spin-splitting energy 176.76 meV. The separation indicates that the CKL slab has spin polarization, which is a strong correlated effect stimulated by the correlated interaction of spin-up and spin-down electrons. The average magnetic moment of each carbon atom is about 0.056 μB, and the magnetic moments of atoms on the triangular rings are larger than those of other atoms.

In addition, the series structures in the CKL family are also a kind of carbon foams which have large specific surface area. Therefore, besides their interesting electronic properties, like the normal carbon foams, they can be also used in some other fields, such as catalysts, ion batteries and energy storage [44-46].

## 4. Conclusions

We proposed a CKL family by using the DFT method and studied their electronic properties. From CKL1 to CKL4, the in-plane bands around the Fermi level gradually change from dispersive bands to flat bands, while in the out-of-plane direction a Dirac point formed in the band structures of CKL3 and CKL4. Therefore, in CKL3 and CKL4, there are coexistence of flat band and Dirac band around the Fermi level, implying that these structures may can be served as superconductors. A tight-binding model is used to explain the evolution of the bands induced by the orbital interactions in the structures. The electronic properties of the CKL slab are also studied. Flat bands and spin-polarization induced by hole doping are observed.

## Acknowledgments

This work was supported by the National Natural Science Foundation of China (Nos. 51176161, 5 1376005 and, 11474243) and the Hunan Provincial Innovation Foundation for Post-graduate (No. CX2015B211).SZ acknowledges support by US DOE under Grant No.DE-SC0002623.

# Supporting Information (SI)

## Content:

1. Geometrical information of CKL1-CK4.
2. The coordinates of the slab.

1. Geometrical information of CKL-CKL4

| System | Space group | Wyckoff position |
|---|---|---|
| CKL1 | $P6_3/mmc$ | 6h(0.4479,0.5521,0.2500) |
| CKL2 | *P6/mmm* | 6m(0.6025,0.3975,-0.5000);6l(0.4520,-0.0960,0.0000) |
| CKL3 | $P6_3/mmc$ | 6h(0.7560,0.3780,0.7500);6h(0.0359,0.5179,0.7500),6h(0.1707,0.5853,0.7500) |
| CKL4 | *P6/mmm* | 6m(0.6323,0.2647,0.5000),6l(0.3952,0.6048,0.0000),6l(0.4469,0.5530,0.0000),6m (0.4740,0.5259,0.5000) |

Table S1. The space group and Wyckoff position of CKL1-CKL4

2. The coordinates of the slab

The coordinates of the slab are provided in the format of the input file of VASP, which is the POSCAR.

```
CKL2-slab
1.0
        8.3313999176         0.0000000000         0.0000000000
       -4.1656999588         7.2152039777         0.0000000000
        0.0000000000         0.0000000000        16.3750000000
    C    H
   18   12
Direct
     0.602169991         0.204339981         0.500000000
     0.397830009         0.795660019         0.500000000
     0.795660019         0.397830009         0.500000000
     0.204339981         0.602169991         0.500000000
     0.602169991         0.397830009         0.500000000
     0.397830009         0.602169991         0.500000000
     0.453020006         0.546980023         0.574949980
     0.546980023         0.453020006         0.425050020
     0.453020006         0.906040013         0.574949980
     0.546980023         0.093959987         0.425050020
     0.093959987         0.546980023         0.574949980
     0.906040013         0.453020006         0.425050020
     0.546980023         0.453020006         0.574949980
     0.453020006         0.546980023         0.425050020
     0.546980023         0.093959987         0.574949980
     0.453020006         0.906040013         0.425050020
     0.906040013         0.453020006         0.574949980
     0.093959987         0.546980023         0.425050020
     0.418900013         0.581099987         0.634199977
     0.581099987         0.418900013         0.365800023
     0.418900013         0.837800026         0.634199977
     0.581099987         0.162199974         0.365800023
     0.162199974         0.581099987         0.634199977
     0.837800026         0.418900013         0.365800023
     0.581099987         0.418900013         0.634199977
     0.418900013         0.581099987         0.365800023
```

| | | |
|---|---|---|
| 0.581099987 | 0.162199974 | 0.634199977 |
| 0.418900013 | 0.837800026 | 0.365800023 |
| 0.837800026 | 0.418900013 | 0.634199977 |
| 0.162199974 | 0.581099987 | 0.365800023 |